\documentstyle[11pt,dunk2001_asp,twoside,psfig]{article}
\def\edcomment#1{\iffalse\marginpar{\raggedright\sl#1\/}\else\relax\fi}
\reversemarginpar

\begin{document}
\title{New Results on Bar-Halo Interactions}
\author{E. Athanassoula}
\affil{Observatoire de Marseille, 2 place Le Verrier, 13248 Marseille
  cedex 04, France}

\begin{abstract}
In this paper I argue that, far from necessarily hindering bar
formation in disc galaxies, inner haloes may stimulate
it. This constitutes a new instability mechanism by which bars can grow. 
To show this I use a number of $N$-body simulations whose initial 
conditions have identical discs and more or less concentrated
haloes. They
show that the bar that grows in the more halo-dominated environment is
considerably stronger than the bar that grows in the more disc-dominated
environment. This result is obtained from simulations with live haloes, 
i.e. composed of particles which 
respond to the disc and take part in the evolution. On the other
hand, if the halo is rigid, it hinders or quenches bar formation, as
expected. Comparison of two simulations which are identical in
everything, except that the halo is live in the first one and rigid in
the second one, leads me to suggest that the halo response can help the
bar grow. Following the orbits of the stars in the halo, I find that
a considerable fraction of the halo particles are in resonance with
the bar. The halo may thus take angular momentum from the bar and
stimulate its growth. I finally discuss whether and how the results of
the $N$-body simulations can be applied to real galaxies.  
\end{abstract}

\section{Introduction}

It is by now well established that galactic discs can be bar unstable
(e.g. Miller, Prendergast \& Quirk 1970; Hohl 1971). In the
quest for stability three main stabilising mechanisms 
have been proposed (see e.g. reviews by
Athanassoula 1984, Sellwood \& Wilkinson 1993 and references
in either) : 

i) The disc could be immersed in a massive spheroid, e.g. a
bulge and/or an inner halo (Ostriker \& Peebles 1973). 

ii) The disc could be hot or have a hot center (Athanassoula 1983,
Athanassoula \& Sellwood 1986). 

iii) The galaxy could be sufficiently centrally concentrated to  stop
the initially linear wave from reaching the center (Toomre 1981; Sellwood \&
Evans 2001). 

Each one of these mechanisms has generated a lot of discussion,
both regarding its efficiency and the way it can be applied to real
galaxies. Here I will address the first of them, which was also
historically the first to be introduced, and show that, contrary to
what has been argued so far, inner haloes can, at least in some cases,
enhance the bar. For this I will use a series of numerical simulations
of disc-halo systems, described by Athanassoula \& Misiriotis
(2002, hereafter AM02) and Athanassoula (2002b in preparation, hereafter
A02b). I describe the
simulations and their results in section 2, and in
section 3 I discuss the role of the halo. Finally in
section 4 I give a general discussion and address
the applicability of my results to real galaxies.

\section{Simulations of and results on bar formation}
\label{sec:simul}

\begin{figure} 
\centerline{\vbox{
\psfig{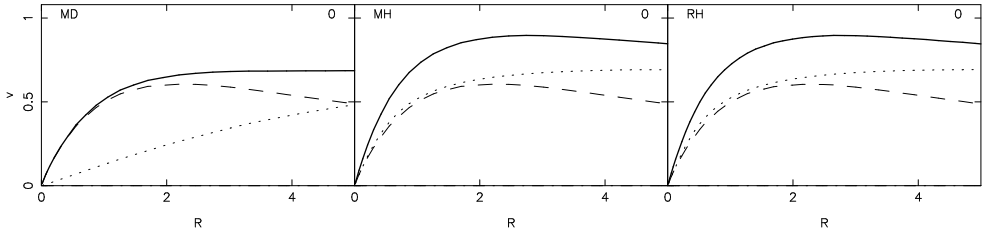}}}
\caption{Circular velocity curves of the three models discussed
    in section~2 (solid lines). The contribution of the
    disc component is 
    given by a dashed line, and that of the halo by a dotted line. The
    left panel corresponds to simulation MD, the middle one to MH and
    the right one to RH.}
\label{fig:inrotcur}
\end{figure}

\begin{figure} 
\centerline{\vbox{
\psfig{{figure=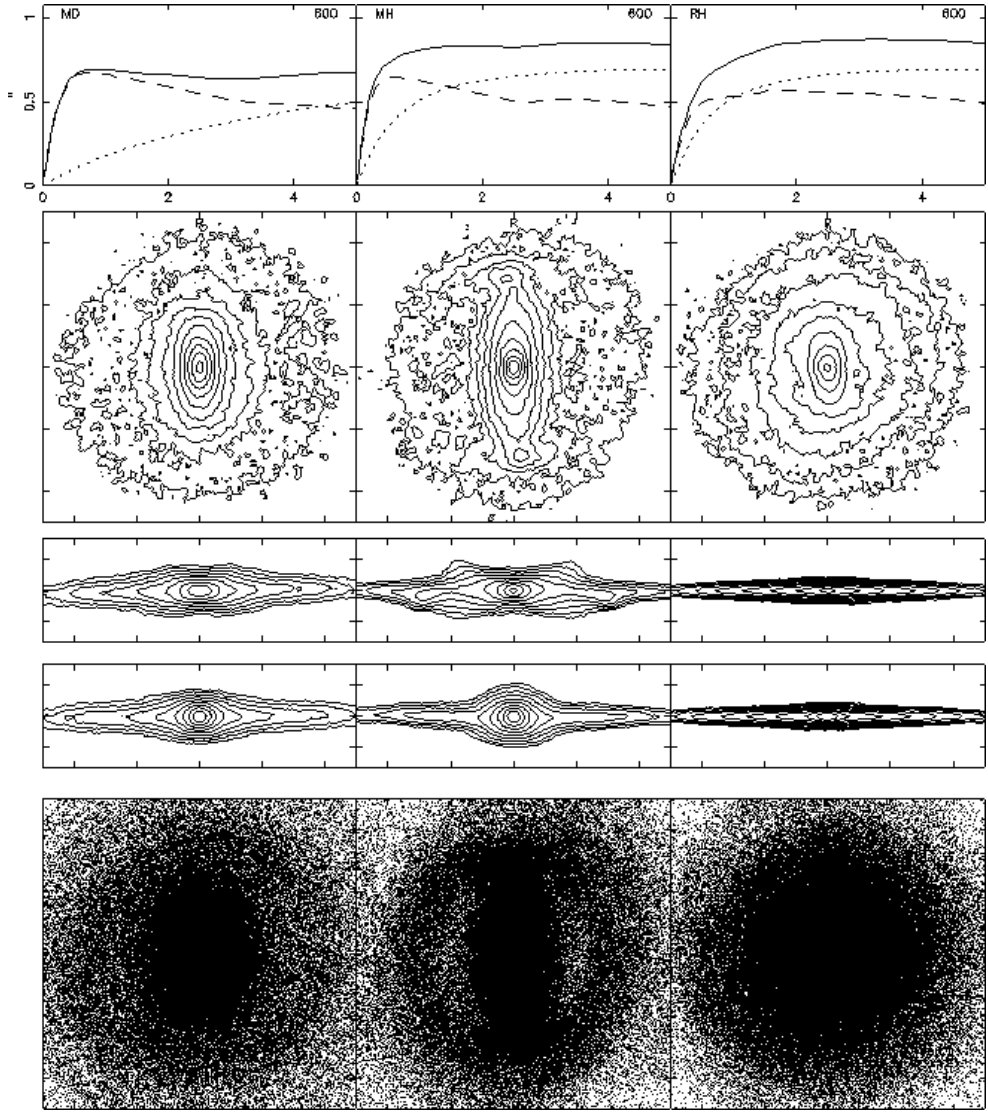,width=13cm}}}}
\caption{Results of the three simulations at t=600. The left
    panels correspond to model MD, the middle ones to model MH and the
    right ones to model RH. The upper panels give the ``rotation
    curves'' (solid lines), together with the contribution of the disc
    (dashed lines) and halo (dotted lines) components. The next three
    rows give the isodensities of the disc component when seen
    face-on (second row) and edge-on (third and fourth row). In the
    third row the bar is see side-on and in the fourth row end-on. 
    The last row gives again the face-on view of the disc
    component, but now as a dot plot. The size of the square box 
    for the second and fifth row of panels is 10 initial disc scale 
    lengths, and all panels, except for the upper row, have the 
    same linear scale.}
\label{fig:basic}
\end{figure}

I will present here the results of three $N$-body simulations made
with the GRAPE-5 systems of Marseille observatory (Kawai et
al. 2000). The three initial conditions have identical exponential
discs of unit mass and scale length, and $Q$ = 1.2. Their halos are five
times as massive as their discs and have different degrees of central
concentration. They are initially thermalised, spherical and
non-rotating. The halo of the
first simulation has initially a mass of 0.4 within three disc scale lengths,
while the halo of the other two has a mass of 1.4 within that
distance. The initial circular velocity curves (loosely called
hereafter rotation curves) of the three models are shown in
Fig.~\ref{fig:inrotcur}. In the first simulation, hereafter MD, the
disc dominates in the inner parts, roughly up to $r$ = 5, whereas in
the other two (hereafter 
MH and RH, respectively), the halo contribution is slightly larger 
than that of the disc up to the maximum of the disc rotation curve, and
considerably so after that. Model RH is identical to model MH,
except that its halo is rigid -- i.e. given by a potential imposed
on the disc particles -- and does not evolve during the simulation. 
Models MD and MH are similar to the ones described in AM02, but have somewhat
thinner discs. Details on the initial conditions and on the
simulations, as well as results for other such simulations, 
can be found in AM02 and in A02b. A reasonable calibration (AM02) 
gives that $t$ = 500 corresponds to 7 $\times 10^9$ years.  
Different values, however, can be obtained for different scalings of 
the disc mass and scale length.

The results of the evolution at $t$ = 600 are summarised in
Fig.~\ref{fig:basic}. The upper row of panels gives the three ``rotation
curves''. We note that the disc material has moved
considerably inwards in the two first cases, but much less so
in the third one. In model MD the difference between $t$ = 0 
and $t$ = 600 is
quantitative, since the inner parts were initially disc dominated and
are still so after the evolution, albeit to a larger degree. On 
the other hand for model MH the difference 
is qualitative, since the inner regions are now dominated by
the disc, contrary to what was the case initially. The remaining
panels give the face-on, side-on and end-on views of the disc/bar
component. The differences between the three models are striking.
Comparing models MD and MH we note that {\it the bar that grew
  in the initially more halo dominated environment is stronger 
  than the bar that
  grew in the disc dominated environment} (AM02). It is longer and its
isophotes are more rectangular-like (AM02). Comparing the Fourier components,
the surface density profiles and the isophotal shapes of the $N$-body
bars with those of
observed barred galaxies, I find that the MD-type bars have observed
properties reminiscent of those of late-type bars, while
MH-type bars resemble early-type strongly barred systems (Athanassoula 2002a). 

Very strong differences can also be found when we compare model MH
with model RH. Model MH has a very strong bar, while model RH has a
mild oval distortion, or a very weak bar. The difference between the
lengths of the two bars is striking -- the MH bar being 2 or 3 times
longer that the RH ``bar''. Their edge-on views are also very
different. Model MH seen side-on has a strong peanut or X-shape, and a
big bulge-like protuberance if seen end-on. Model RH shows no such
features. 

The difference between the three bars is also clear when I plot the
relative Fourier component of the face-on distributions of the discs 
(e.g. AM02). The maximum values of the 
relative $m$ = 2 component for models 
MH, MD and RH at time 800 are 0.68, 0.45 and 0.23, respectively.

The three models differ also in the way their bars evolve. The pattern
speed of model MH starts off higher than that of model MD, but ends
considerably smaller. The pattern speed of MD also decreases with
time, but much less so. In this our results agree at least qualitatively
with those of Debattista \& Sellwood (1998) obtained for different
galactic models and with a different type of code. A more quantitative 
comparison, including the time evolution of the ratio of the
corotation radius to 
the bar length, is underway. The pattern speed of the ``bar'' in 
RH  can not be measured reliably before $t$ = 300, and after 
that does not show any signs of decrease. For models MD and MH
there is exchange of energy and
angular momentum between the disc and halo components, so that the
halo, which was initially non-rotating, displays rotation after the
bar has grown. This is small for model MD and considerable for model
MH, for which shortly after $t$ = 800 the halo has roughly a third of
the angular momentum of the disc, i.e. a quarter of the total.
 
\section{The role of the halo}
\label{sec:halo}

\begin{figure} 
\plotfiddle{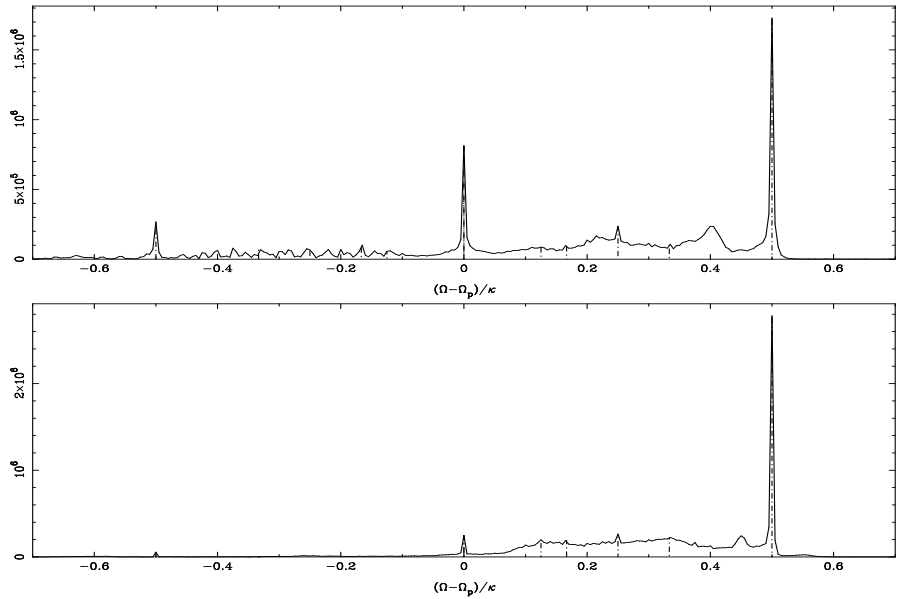}{7.9cm}{270}{50}{40}{-200}{250}
\caption{Number density of disc particles as a function of the 
frequency ratio $(\Omega -
\Omega_p) / \kappa$, for simulation MD (upper panel) and MH (lower
panel) at $t$ = 800. The dot-dashed vertical lines give the positions of the 
main resonances.
}
\label{fig:disc_res}
\end{figure}

\begin{figure} 
\plotfiddle{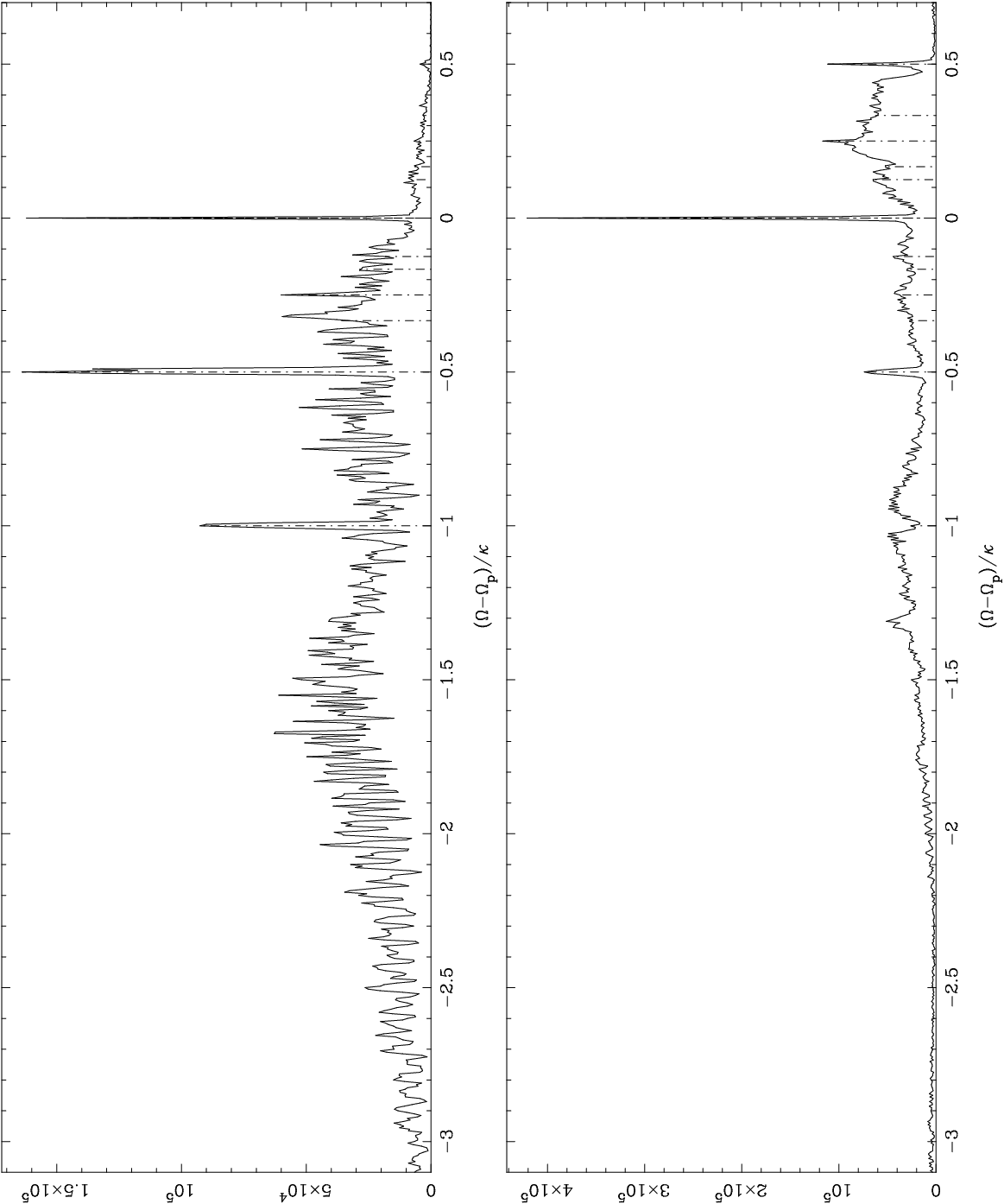}{7.9cm}{270}{50}{40}{-200}{250}
\caption{Number density of halo particles as a function of the
frequency ratio $(\Omega -
\Omega_p) / \kappa$, for simulation MD (upper panel) and MH (lower
panel) at $t$ = 800. The dot-dashed vertical lines give the positions of the 
main resonances. 
}
\label{fig:halo_res}
\end{figure}

What is the role of the halo in these simulations, and why is it that
the discs which are immersed in more massive {\it live} haloes make stronger
bars? Although I still do not have a full quantitative answer to this 
question, I
will present here results which can bring useful qualitative understanding. 

I have always found that understanding the orbits in a given system is
a crucial step towards understanding its dynamics. In particular
Lynden-Bell \& Kalnajs (1972) have shown that stars at resonances can
absorb or emit angular momentum, thus driving the evolution of the
disc. I therefore froze the
potential at a few selected times during each simulation and followed
the orbits 
of 100\,000 disc particles, and of an equal number of halo ones. From
these I calculated the basic frequencies of each orbit, namely
$\Omega$, $\kappa$ and $\kappa_z$. Here $\Omega$ is the angular
frequency, $\kappa$ is the epicyclic frequency and $\kappa_z$ is the
vertical frequency. An orbit is resonant if there are three integers
$l$, $m$ and $n$ such that 
$l \kappa + m \Omega + n \kappa_z = m \Omega_p$, where $\Omega_p$ is 
the pattern speed of the bar.

The main resonances occur for small integer numbers. I will here
restrict myself to radial (planar) resonances, for which $n$ = 0. The most
important such resonances are the inner
Lindblad resonance (ILR), where $l$ = -1 and $m$ = 2, the corotation
resonance (CR), where $l$ = 0, the outer Lindblad resonance (OLR),
where $l$ = 1 and $m$ = 2, and the inner and outer
ultra-harmonic resonances (IUHR and OUHR respectively), for which $m$ =
4 and $l$ = -1 and 1 respectively. At corotation resonance the
particles have the same angular frequency as the bar, while in the
other resonances they make $m$ radial oscillations in the time they
make $l$ revolutions around the center of the galaxy. Calculating the
basic frequencies, and in 
particular $\Omega$, is far from straightforward. In fact I am still
working towards an optimum method, combining reliability, robustness
and speed, so that the results given here can be considered as
somewhat preliminary. Nevertheless, they are sufficiently reliable for
80\% to 90\% percent of the particles and so they warrant discussion.

Figures~\ref{fig:disc_res} and \ref{fig:halo_res} show the 
number density of particles/orbits that have 
a given value of $(\Omega - \Omega_p) / \kappa$ as a function of this
ratio. Let me first describe the results for the disc components. 
The distribution is far from homogeneous and there are
strong peaks at the location of the main resonances. The highest peak,
both for the MH and the MD disc, is for the 
ILR, where $(\Omega - \Omega_p) / \kappa$ = 0.5. Indeed orbits
making two radial oscillations in the time they make one revolution
around the center of the galaxy are the backbone of the bar. The peak
is higher, by roughly 60 \%, in the MH case than in the MD case, in
agreement with the fact that the bar in this model is stronger
(cf. Fig.~\ref{fig:basic}). The height of the CR peak is sizeable for
model MD, and small for MH. The ratio of the height of the CR peak to that
of the ILR is 0.47 for the MD case, while for model MH it is only
0.09. Model MD has a sizeable peak also for $(\Omega - \Omega_p) /
\kappa$ = -0.5, i.e. at the OLR. The
differences between the two models are due to their different
corotation radii. Thus at time 800 the CR radius is 6.7 for model
MH and 3.3 for model MD. Similar differences are found for the
OLR radii of the two models. Therefore these two resonances are in the
outer parts of the MH disc and can trap only few particles. This
is not the case for model MD, and the differences in the trappings are
reflected in the differences in the heights of the respective resonant peaks. 

The big surprise, however, comes from the halo component. So far
considered as non-, or little, responsive, it shows, on the contrary,
unmistakable signs of strong resonances with the bar. Since we have
analysed the same number of 
particles from the two components, while there are roughly five times
more halo than disc particles, the halo numbers should be multiplied by
roughly five, thus highlighting the importance of the halo for the
dynamical evolution of the galaxy. 

There are important differences between the orbital structures of the MH and MD
haloes, as was the case for the corresponding discs. The ILR peak of
model MH is relatively high (more than 20 times higher than the
corresponding MD peak) and there is considerably more material between
ILR and CR than in model MD, while the CR peak in model MH is more
than twice as high as 
the corresponding peak in model MD. All these are in agreement with the fact
that the MH halo is much more concentrated than the MD one. Further
out the situation shifts. The OLR peak of the MD model is more than twice
as high as the MH one, and the -1:1 peak is also very high in MD,
while for model MH this resonance has been depleted of its 
particles. The amplitude of the CR peak for model MD shows
considerable evolution with time, nearly doubling from 500 to 800. The
role of these differences in the dynamical evolution of 
the galaxy will be discussed elsewhere.

Halo stars in resonance with the bar can exchange
energy and angular momentum with it and thus influence its
evolution (Tremaine \& Weinberg 1984). In general,
the halo resonances will absorb
angular momentum. Since the bar is inside corotation, it has negative
energy and angular momentum, and thus the effect of halo resonant
stars will be to destabilise it, i.e. will lead to stronger bars. This
is in good agreement with the results on angular momentum transfer
discussed at the end of the last section. Since the disc, the bar and
the final halo component rotate in the same direction, 
the halo will take positive angular momentum from the disc/bar
component and thus will further destabilise the bar.

I am thus proposing a new instability mechanism, by which the
halo will stimulate bar growth. This of course will only work
if the halo is non-rigid and is capable of absorbing positive angular
momentum. Similarly the bar should also be non-rigid. Further work
quantifying this mechanism and studying its  efficiency in
different types of models and in real galaxies is in progress (A02b). An
analytical description of this instability, including the effect of
the non-planar resonances, and a comparison with the $N$-body results 
is forthcoming, in collaboration with M. Tagger and F. Masset.

\section{Discussion}
\label{sec:discuss}

In the above I have compared three simulations starting off with
identical discs, but different halo components. Any differences in
their dynamical evolution should thus be attributed to the haloes.
The strongest bar forms in the most halo dominated case, provided this
is live, followed by the one in the disc-dominated case. In the
simulation with the rigid halo there is only a very weak bar, or mild oval
distortion, in the inner part. I thus reach the interesting conclusion
that haloes can, at 
least in some cases, stimulate the bar instability and lead to stronger
bars. This can be understood by a frequency analysis of the halo
orbits, which reveals a large number of resonant orbits. Since these
can exchange energy and angular momentum with stars at other
resonances (Lynden-Bell \& Kalnajs 1972) they can stimulate the bar
instability, contrary to previous beliefs. 

The evolution of the galaxy leads to considerable concentration of the
disc material in the central areas. Thus model MD starts off as disc
dominated in the central parts, and, with time, the disc further enhances its
superiority. Model MH starts off quite differently. Initially the 
halo is slightly more important than the disc within the radius at 
which the disc rotation curve is maximum, and considerably more so 
at larger radii, as witnessed from its circular velocity curve, 
shown in Fig.~\ref{fig:inrotcur}. 
The central concentration of the disc increases considerably with time, 
so that, after the bar has grown,
the disc dominates in the inner region. This may contribute an 
additional argument to the long standing debate of
whether galactic discs are maximum or sub-maximum (e.g. Athanassoula, 
Bosma \& Papaioannou 1987; Bosma 1999, 2000;
Bottema 1993; Courteau \& Rix 1999;
Kranz, Slyz \& Rix 2001; Sellwood 1999, Weiner, Sellwood \& Williams 2001).  

Sackett (1997) and Bosma (2000) give a simple working definition 
to distinguish 
between maximum and sub-maximum discs, based on the value of 
$\gamma = V_{d,max} / V_{tot}$, where $V_{d,max}$ is the circular 
velocity due to the disc component 
and $V_{tot}$ is the total velocity, both calculated at a radius 
equal to 2.2 disc scale lengths.  
According to Sackett (1997) this ratio has to be at least 0.75 for 
the disc to be considered maximum or maximal. In the simulations 
it is not easy to define a disc scale length after the bar has 
formed, so I will calculate $\gamma$ at the radius at which the disc 
rotation curve is maximum, which is a well defined radius and 
is roughly equal to 2.2 disc scale lengths in the case of an 
exponential disc. Model MD starts off with $\gamma > 0.75$, so that
the disc starts maximum and stays so all through the simulation.
In fact the value of $V_{d,max} / V_{tot}$ increases somewhat 
with time.  Model MH has initially a value of $\gamma$ around 0.68,
i.e. close
to the value of 0.63 advocated by Bottema (1993), and is therefore
initially sub-maximum. This value, however, 
increases abruptly after the bar has formed, so that the 
disc can be considered maximum well before the time shown in 
Fig.~\ref{fig:basic}, with a value of $\gamma$ roughly equal to 0.86. 
Thus the formation of the bar leads the disc to evolve from sub-maximum 
to maximum, and hence strongly argues for maximum discs in disc 
galaxies with strong bars. This means that if we observe a strong bar 
in a disc galaxy the above simulations argue strongly and quantitatively 
that the underlying disc is maximum. The existence of gas should
not alter this result. Indeed if the gas leads to a density distribution with 
a weak bar or no bar, then the above argument will be irrelevant, 
since it applies only to galaxies with strong observed bars. On 
the other hand, if the resulting bar is strong, then it should have 
rearranged the disc material sufficiently for the above argument to hold. 

We can reach similar results about the disc-to-halo mass ratio
if we use the criterion of Athanassoula, Bosma \& Papaioannou (1987),
who examined what spiral perturbations can grow in a given 
disc/halo decomposition of an observed rotation curve. In a similar way 
I can calculate the $m$ component that
will be strongest amplified via the swing amplification mechanism 
(Toomre 1981) in my simulations at or around the radius at which 
the disc rotation curve 
reaches its maximum. I find that, for model MD, it is the 
$m$ = 2 all through the simulation, as
expected. The initial disc for model MH is certainly not maximum. I
find that at $t$ = 0 higher $m$ components will be the most strongly
amplified. The evolution, however, changes this, so that after the bar 
has grown it is the $m$ = 2 component that is the strongest amplified
at or around the radius at which the disc rotation curve 
reaches its maximum. Both MD and MH models thus have, after the bar
has grown, a disc which is intermediate
between the ``no $m$ = 1'' and ``no $m$ = 2'' limits advocated by
Athanassoula, Bosma \& Papaioannou (1987) for real galaxies. 
Indeed these authors made a
link between the structure present in a disc at a given time and the
underlying halo mass at that time (not the initial halo mass) and thus
their results are in good agreement with the above simulations, and 
many other similar ones (e.g. AM02 and A02b) .

Apart from the halo, several other parameters can influence the
formation of a bar. In particular let me stress the importance of
the velocity dispersion of the disc particles, the effect of gas, 
and the effect of
the velocity and mass distribution in the halo, as well as the
existence of a gaseous companion. A complete description of all these
effects is well beyond the scope of the present contribution and
will be presented elsewhere. Let me
just add a few preliminary words about the effect of the disc velocity
dispersion. A sequence of MH-type galaxies shows that for larger
initial velocity dispersion of the disc particles the bar is less
strong and for sufficiently high values, becomes oval, or quasi-circular, in
good agreement with what was found for 2D models (Athanassoula 1983). 
A sequence of MD-type models is more complicated. In these models the
bar grows faster and becomes very long and strong. At that time,
however, a strong buckling instability develops which leads to a
considerable decrease of the bar amplitude. The final amplitude of the
bar is a result of the competition between these two effects and this
may be close. Only at sufficiently large values of the velocity 
dispersion can we be sure that the resulting oval will be very
thick, as in the MH sequence. All the above are rather preliminary and
will be discussed at length elsewhere.

Finally the mass and velocity distribution of the halo component,
together with the bar pattern speed and its time dependence, should 
influence how each of the resonance regions is populated and how 
responsive it is, and therefore influence its ability to exchange 
energy and angular momentum with the bar. 
Since very little is known on the composition of the halo, let alone
about the distribution of the matter in it, it is very difficult to
pursue this issue further. Nevertheless the arguments in
section~3 lead to the prediction that at least some of
the stars of the visible halo should be in resonance, in as much as they
trace the relatively inner parts of the halo. Testing this would
necessitate accurate information on the six phase space coordinates of
a sufficiently large number of halo stars, as well as a sufficiently
accurate description of the halo potential and the bar pattern
speed. Our own Galaxy, which is barred, is the only place where 
advances with future astrometric satellites may make this possible, if we
concentrate in areas which could have a high fraction of resonant stars.

\acknowledgments

I would like to thank Albert Bosma for many discussions on barred
galaxies, M. Tagger for discussions on the description and effect of
the halo resonances, A. Misiriotis for his collaboration on the software
calculating the orbital frequencies and J.~C. Lambert for his help
with the GRAPE 
software and the administration of the simulations. I
would also like to thank the Region PACA, the
INSU/CNRS, the University of Aix-Marseille I and the IGRAP for funds to develop
the GRAPE computing facilities used for the simulations discussed in
this paper.

\section*{Discussion}

\noindent {\it Kormendy:\, } You concluded that disks of barred
  galaxies are maximal. Ortwin Gerhard concluded in the previous talk
  that the disk of our Galaxy is close to
  maximal. But Piet van der Kruit concluded that the typical late-type
  disk is substantially sub-maximal. This is a microcosm of the debate
  about maximal versus sub-maximal disks that has been going on in the
  literature for a long time. I wonder what we can learn about this
  apparent disagreement and in particular whether we can make
  progress in resolving it because we have proponents of both points
  of view at this meeting.\\

\noindent {\it Athanassoula:\, } My simulations show that even discs
which are initially sub-maximum can form bars and thus evolve to
maximum. They thus argue that discs with strong bars are maximum,
or near so. Concerning non-barred galaxies there are, as you just said,
results arguing for maximum discs and others arguing
for sub-maximum discs. I thoroughly agree that a debate on this
subject would be most useful and I would like to mention that it will
be the topic of a panel discussion in the 6th Guillermo Haro meeting,
which will take place in Puebla (Mexico) in November 2001.\\

\noindent {\it Illingworth:\, } What are the initial conditions
  for your models? I am wondering how you would relate these to actual
  galaxies. In a sense, how realistic are they compared to how
  galaxies are built up?\\

\noindent {\it Athanassoula:\, } The instability mechanism which 
I proposed here,
namely stimulating the bar growth by its interaction with the
halo, depends on the present day properties of the halo and the bar, and only
indirectly on the formation history of the galaxy. Since the
properties of my $N$-body bars are in good agreement with those of
bars observed at $z$ = 0, I do not have to worry overly about 
how the galaxy was built up, at least in as much as the physics 
of the mechanism I proposed here is
concerned. Of course if we were able to follow the formation of discs
and bars starting from initial conditions resembling those of
galaxies observed at high $z$, including a fair fraction of gas
and non-equilibrium initial conditions, we would be able to learn
a lot about disc galaxy formation and evolution.\\

\end{document}